\documentclass[12pt]{article}
\date{}
\begin{document}

{\bf Comment on Hess-Philipp anti-Bell and 
Gill-Weihs-Zeilinger-Zukowski anti-Hess-Philipp
arguments}

\medskip

Andrei Khrennikov

\medskip

{\it International Center for Mathematical
Modeling}

{\it in Physics and Cognitive Sciences,}

{\it  University of V\"axj\"o, S-35195, Sweden}\footnote{Email:Andrei.Khrennikov@msi.vxu.se}

\medskip

\begin{abstract} We present comparative analysis of  
Gill-Weihs-Zeilinger-Zukowski arguments directed against
Hess-Philipp anti-Bell arguments. In general we support
Hess-Philipp viewpoint to sequence of measurements in the EPR-Bohm
experiments as stochastic time-like process. On the other hand, 
we support Gill-Weihs-Zeilinger-Zukowski arguments against
the use of time-like correlations as the factor blocking the  derivation
of Bell-type inequalities. We presented our own time-analysis of 
measurements in the EPR-Bohm experiments based on the frequency 
approach to probability. Our analysis gives strong arguments in 
favour of local realism. Moreover, our frequency analysis 
supports the original EPR-idea that quantum mechnaics is not complete.
\end{abstract}

\medskip

Last two years K. Hess and W. Philipp published the series of papers
in that they presented probabilistic arguments against the 
use of   Bell's inequality [1] as the crucial argument
against  local realism, see [2]. On the other hand, 
R. Gill, G. Weihs, A. Zeilinger, M. Zukowski published preprint 
[3] containing rigid critique of Hess-Philipp anti-Bell considerations.
We would like to analyse arguments presented by both sides as well as present
our own probabilistic analysis of Bell's framework. This analysis is based on
the {\it frequency approach} to probability theory (as an alternative to the 
standard measure-theoretical approach).

As it was rightly mentioned by authors of [3],
publications of K. Hess and W. Philipp [2] have really drawn a lot of attention,
especially since the work featured on Nature's web pages. Therefore it is very 
important to have the correct understanding of the role of these works in
the study of Bell's arguments against local realism. First 
it should be noticed that investigations [2] 
are not at all first works containing probabilistic analysis of Bell's 
arguments, see e.g. [4]-[7], see also book [8] on the extended bibliography.
Moreover, we remark that (it may be just by chance) in all these papers, besides
of one exception, namely the paper of R. Gill [7], there were obtained anti-Bell
conclusions supporting  local realism. We do not plan even to try to 
consider this huge series of probabilistic anti-Bell publications to compare with works [2].
We shall just point out main anti-Bell probabilistic arguments:

{\bf Contextualism.} The probability distribution ${\bf \rho}(\lambda)$
of hidden variables could not be chosen independently of
experimental settings. In fact, instead of the distribution ${\bf \rho}(\lambda)$
(as Bell did), we have to consider distributions
${\bf \rho}_{a, b^\prime}(\omega)$ corresponding to experimental settings
$a, b^\prime.$ Such a dependence appears quite naturally if we assume
that the general hidden variable, HV, $\omega$ contains not only the  HV 
$\lambda$ describing the state of correlated particles, but also 
HVs $\omega_a$ and $\omega_{b^\prime}$ corresponding to states 
of Stern-Gerlach magnets with settings $a$ and $b^\prime.$ Thus, in fact,
HV has the form $\omega=(\omega_a, \lambda, \omega_{b^\prime}).$ This is
very natural experimental assumption (see my further frequency 
analysis of the EPR-Bohm experiment). If we
follow to contextualism it would be impossible to use Bell's representation 
of probabilities:
\begin{equation}
\label{B}
{\bf P} (A,B^\prime) = \int {\bf \rho}(d \lambda) {\bf P} (A,B^\prime/ a,b^\prime, \lambda).
\end{equation}
We have to use the general representation:
\begin{equation}
\label{BB}
{\bf P} (A,B^\prime) = \int {\bf \rho}_{a, b^\prime}(d\omega) {\bf P} (A,B^\prime/ a,b^\prime, \omega)).
\end{equation}
This blocks the derivation of Bell-type inequalities.

{\bf No factorisation.} The Bell-Clauser-Horne factorability condition (sometimes called locality condition):
\begin{equation}
\label{BF}
{\bf P} (A,B^\prime/ a,b^\prime, \lambda)={\bf P} (A/ a, \lambda){\bf P} (B^\prime/b^\prime, \lambda)
\end{equation}
does not hold true. If it is the case, then derivations of  Bell-type inequalities
are blocked.\footnote{We remark that (\ref{BF}), in fact, has nothing to do 
with locality. This is purely probabilistic condition of independence.
Nevertheless, it is not so easy to motivate the violation of (\ref{BF}) in the EPR-framework.
We shall discuss this problem (in fact, the problem of the right understanding of
the statistical independence) in our frequency analysis of the EPR-Bohm experiment.}

{\bf No probability distribution of hidden variables.} It is clear that if probability
distributions (measures) of HV  do not exist at all, then the derivations of 
Bell-type inequalities
are blocked.\footnote{Of course, the reader educated in the framework of the conventional
measure-theoretical probability theory (Kolmogorov's axiomatics, 1933) should be
surprised. For him, if there is no measure, then there is no probability at all.
However, if we leave the domain of Kolmogorov's probability theory and consider
e.g. frequency probabilistic model  (that is essentially more natural from
the experimental viewpoint), then we can find that relative frequencies of observed
quantities could stabilize in the absence of the measure-probability for HV, see
examples in book [8].}

We notice that we can escape the problem related to contextualism if
assume that the simultaneous probability distribution ${\bf \rho}_{a, b^\prime}(d\omega)$
can be factorized producing, finally, the Bell-Clauser-Horne factorability condition.

{\bf Hess-Philipp's contextual probabilistic model.} We turn back to works [2].
 As we have noticed, we are not interested
in the detailed analysis. It is evident that works [2] belong to the class
of contextual investigations, see e.g. :

{\small "The essence of our approach is the introduction of setting and station specific
time-like parameters as well as time related setting dependent parameters 
$\lambda_a^\star$ on one side and $\lambda_b^{\star\star}$ on the other, 
that codetermine the functions $A,B$ in addition to the correlated source
parameters $\lambda.$... We also show that these parameters lead in a natural way to setting 
dependent probability measures for the parameters without spooky action at a distance,"
[2].}

If we do not pay attention to "time-considerations" (see further
analysis of this crucial factor) and restrict us to purely probabilistic 
considerations, then we simply get a new contextual probabilistic model, Hess-Philipp's
model. Moreover, the Bell-Clauser-Horne factorability condition is also violated in Hess-Philipp's
model:

{\small "Bell type proofs permit any number and form of parameters as long 
separate integrations can be performed over the respective densities i.e. if the joint conditional 
densities equal the product of the individual conditional densities. The introduction
of time-like parameters presents then a critical problem since other parameters in the argument of
the functions $A$ and $B$ may depend on time," [2]}

So, from the first sight Hess-Philipp's model is just one of contextual models with the violation
of the Bell-Clauser-Horne factorability condition. It is clear that in such a model Bell's inequality 
could be violated. However,
K. Hess and W. Philipp not only manipulate with contextual probabilistic nonfactorizable
distributions,
but they also provided the justification of the appearance of such distributions
by using time-like  parameters. And this justification is the crucial point of their
investigations.

{\bf The role of time in the EPR-Bohm measurements and Hess-Philipp's  model.}
The crucial point of Hess-Philipp's considerations is the recognition of the role
of time in the appearence of contextual probabilities and the violation of 
the Bell-Clauser-Horne factorability condition. Hess and Philipp correctly 
observed that the sequence of measurements produces a time-like stochastic 
process. This time structure should be in some way incorporated in the structure of the 
probability space used to describe these measurements. We could not operate
with abstract probability measures as J. Bell and many others, e.g. 
the authors of [3], did. In particular,
by taking into account  the time-structure of the process K. Hess and W. Philipp
demonstrated that the Bell-Clauser-Horne factorability condition does not look
so natural.  It could be violated without inducing nonlocality.
I think that in general it is the right conclusion.

However, the concrete mechanics [2] inducing this violation does not look 
so natural and justified in [2]. In [2] there is promoted the point of view
that the crucial role is played by time-correlations between processes in
measurement devices:

{\small "The core of this demonstration is the mathematical fact that parameters in two stations may be 
conditionally dependent (e.g. during  certain time periods) and simultaneously independent when
no condition are imposed." ... " For example, for time periods during which certain time operators are at work
and/or certain parameters $\lambda$ are emitted from the source, the parameters in station
$S_1$ may be correlated to those in station $S_2$ in other words are conditionally dependent."...
"Our station parameters are correlated by clock time ...", [2].}

In our frequency analysis we shall present another point of view to the source of violation of 
the Bell-Clauser-Horne factorability condition. We also consider a sequence of measurements
as time-process (by using von Mises theory of collectives). Analysis of this process also
implies that the Bell-Clauser-Horne factorability condition should be violated. However, our arguments
differ strongly from Hess-Philipp's arguments. Our frequency analysis demonstrated 
that the Bell-Clauser-Horne factorability condition is violated due to the EPR-correlations.
However, before to present our frequency analysis of the EPR-Bell arguments, we would like to 
turn back to [3].

{\bf The choice which an experimenter is free to make in the laboratory.} 
The authors of the paper [3] wrote about Hess-Philipp's investigation:
"They (Hess and Philipp) claim that time variables should be included in the proof but as 
we show below this is untrue." I could not totally agree with this. Well, considerations
presented in [3] (see also paper [7]  on martingal analysis of the EPR-Bell
arguments) give the strong arguments
against the idea of Hess and Philipp on time correlations between stations $S_1$ and $S_2.$ However,
I think that the general idea on the crucial role of time in this framework was not discredit in 
any way by R. Gill, G. Weihs, A. Zeilinger, M. Zukowski arguments. I think that the main
problem is that these authors are victims of the convectional formalism of probability theory
in that all considerations are performed on the basis of the abstract Kolmogorov
probability space. 

We also remark that the authors of [3]
did strategic mistake in the presentation of anti-Hess-Philipp arguments.
In fact, their
arguments [3] are merely  directed to support classical Bell's arguments against local
realism. In particular, the work [3] contains some proof of Bell's inequality, see also [7].
The role
of this proof in the discussion with  K. Hess and W. Philipp is not clear. In fact,
this is more or less standard proof of Bell's inequality based on the use of {\it counterfactuals,}
see e.g. Stapp  and Eberhard [9]. Such types of proofs were strongly criticized in quantum community.
We would not like go deeply in critical discussions. We simply recall A. Peres:
"Unperformed experiments have no results".  In purely mathematical framework it 
should be noticed that, in fact, this proof
is reduced to the existence of the common Kolmogorov probability space for HV related to different
settings of measurement devices, see
{\bf Contextualism.} This is Bell's original assumption. Following to J. Bell,
the authors of [3] identify this mathematical assumption 
with realism. I think that this is the root of the whole Bell-mystification in quantum theory,
see [8]. 

In fact, we have to differ {\it individual realism} and {\it statistical realism.}
The first one -- the existence of objective properties for individual physical systems.
The second one -- the  existence of simultaneous probabilistic distributions for these
properties. There are no reasons why individual realism should imply statistical realism,
see [8]. 

Thus in [3] it was rightly  pointed out that "the choice which an experimenter is free 
to make in the laboratory" should destroy time-correlations promoted  in [2].
On the other hand, the general critique of the time-random process approach 
[2] could not be considered as totally justified. Moreover, it is not clear
which role plays in [3] the presentation of the old-fashioneds counterfactual proof
of Bell's inequality.

{\bf Frequency analysis of the EPR-Bohm experiment.} Since a few years, I promote 
the use in quantum physics of the frequency approach to probability theory\footnote{Here
probabilities are defined as limits of relative frequencies in long sequences of trials.}
(developed by R. von Mises)
as an alternative to the  measure-theoretical
approach (developed by A. N. Kolmogorov). The main distinguishing feature of the frequency
approach is that here we do not manipulate with abstract probability measures, but 
with concrete random sequences, {\it collectives}, produced in various experiments. Even if we
move later to  probability distributions corresponding to collectives these distributions
would have special structures induced by structures of collectives. In particular, von Mises'
theory is  {\it contextual} by its nature, since probability distributions should depend on 
corresponding collectives.

The detailed frequency analysis of the EPR-Bohm experiment was provided in the preprint [10].
Unfortunately, the work [10] was too mathematical and physicists had large difficulties to follow
to it. In this letter we present the short summary of this frequency 
analysis without to go deeply in mathematical details.\footnote{There is a rather general opinion
that von Mises approach to probability is not justified on the mathematical level of
rigorousness. However, this is not the case. Well, original Mises' definition of randomness
was not mathematically rigorous. However, it was improved, see e.g. [8] for details.
In particular, we can use some restricted classes of place selections (e.g. recursive)
to get mathematically correct theory.} The publication of papers [2] and [3] gives 
the good chance to do this, since there are some similarities in my frequency 
and Hess-Philipp's  time-like 
correlations approaches. The common point of the frequency
and Hess-Philipp's approaches is taking into account the time structure of repetitive  preparation and measurement 
processes in the EPR-Bohm experiments. This gives the possibility to find the internal structure of 
probabilistic measures used to describe HV in EPR-Bohm experiments.

By taking into account the time factor in both approaches we got contextual probabilistic
models without the Bell-Clauser-Horne factorability condition. So the derivation of Bell-like
inequalities is blocked. However, despite the common use of the time structure in both approaches,
sources of correlations are totally different. As well as the authors of [3],
I am quite sceptical to Hess-Philipp's idea on time-like  correlations. The frequency analysis
pays our attention to totally different source of correlations. In fact, this is the original
EPR-source, namely the presence of  (ordinary classical) correlations between 
particles in the EPR-pairs. The crucial point of our analysis is understanding of the fundamental
difference in viewpoints to independence in the conventional Kolmogorov probabilistic model
and the frequency von Mises model. In the conventional model independence is independence of events.
In the frequency model it is independence of collectives. I totally agree with R. von Mises
that the frequency viewpoint to independence is essentially more realistic than the event
approach\footnote{ See e.g.  [8] on discussions and examples demonstrating that "independence-factorisation"
of a probability-measure for events could occur just due to mathematical manipulations with numbers. In general,
such a factorization  has
no physical meaning at all. On the other hand, the notion of independence of two experimental collectives
has the deep physical meaning.}.   

Mathematically the EPR-Bell story looks very simple from the frequency viewpoint:

Let $\lambda_j, j=1,2,...$ be the value of the HV for the $j$th pair of correlated particles
$(\pi^1_j, \pi^2_j)$ produced at the instance of time $t_j=j.$
For  settings $a$ and $b^\prime,$ we consider sequences of pairs
$$
x_{\omega_a, \lambda}= \{ (\omega_{a1}, \lambda_1),...., (\omega_{aN}, \lambda_N),...\}\;,
$$
$$
x_{\omega_{b^\prime}, \lambda}= \{ (\omega_{b^\prime 1}, \lambda_1),...., (\omega_{b^\prime N}, \lambda_N),...\}\;,
$$
and
$$
x_{\omega_a, \lambda, \omega_{b^\prime}}= \{ (\omega_{a1}, \lambda_1, \omega_{b^\prime 1}),...., 
(\omega_{aN}, \lambda_N, \omega_{b^\prime N}),...\}\;,
$$
where $\omega_{aj}$ and $\omega_{b^\prime j}$  are internal states of apparatuses.

The first crucial assumption  that was indirectly  used  by J. Bell and all 
others supporting him (including the authors of [3]) is that all these sequences are
really collectives, i.e. (in particular) limits of relative frequencies in these
sequences exist defining limiting {\bf probability distributions.} Well, there are not so 
many experimental evidences that this is really the case. The only thing that we know
from experiment is that frequencies for results of macroscopic measurements stabilize.
But why should frequencies for internal microscopic states stabilize? In principle,
our macroscopic preparation procedures could violate stabilization of frequencies for HV,
so violate the law of large numbers, see [8] for corresponding examples in that
macro-stabilization is produced by chaotic
micro-fluctuations. In such a case it would be the end of the whole Bell's story.
We again recall that J. Bell and all his adherents mixed {\it individual realism}
and {\it statistical realism.} The absence of probability distributions does not 
imply the absence of objective (may be contextual) properties.

Well, we assume that all considered sequences of HV are really collectives,
so we can freely manipulate with probability distributions used by J. Bell
and others. We come now to the fundamental problem of the whole Bell-story:

{\it Are collectives $x_{\omega_a, \lambda}$ and $x_{\omega_{b^\prime}, \lambda}$
independent?}

If they are really independent, then we get the factorization of
the probability distribution of the collective $x_{\omega_a, \lambda, \omega_{b^\prime}}$
into the product of probability distributions of the
collectives $x_{\omega_a, \lambda}$ and $x_{\omega_{b^\prime}, \lambda}$ and, finally,
the Bell-Clauser-Horne factorability condition
for probabilities. In the opposite case we have no Bell's inequality at all.

However, I do not see any reason why the
collectives $x_{\omega_a, \lambda}$ and $x_{\omega_{b^\prime}, \lambda}$ 
should be independent. Both these collectives contain the same 
correlation-HV $\lambda.$ By the original correlations between 
particles in the EPR-pairs these collectives should be dependent.

{\bf Conclusion.} {\it Frequency analysis of the EPR-Bell 
arguments demonstrated that the Bell-Clauser-Horne factorability condition
should be violated due to dependence of collectives on HV of corresponding 
EPR-pairs. Such  violation of the Bell-Clauser-Horne factorisation condition
has nothing to do with nonlocality. In particular, frequency analysis simply
supports the original EPR-viewpoint that quantum mechanics is  not complete
theory.}

{\bf REFERENCES}

[1] J. S. Bell, {\it Speakable and unspeakable in quantum mechanics.}
Cambridge Univ. Press (1987);
J. F. Clauser , M. A. Horne, A. Shimony, R. A. Holt,
{\it Phys. Rev. Letters,} {\bf 49}, 1804-1806 (1969);
J.F. Clauser ,  A. Shimony,  {\it Rep. Progr. Phys.,}
{\bf 41} 1881-1901 (1978).
 A. Aspect,  J. Dalibard,  G. Roger, 
{\it Phys. Rev. Lett.,} {\bf 49}, 1804-1807 (1982).

[2] Hess K. and Philipp W., {\it Einstein-separability, time related hidden 
parameters for correlated spins, and the theorem of Bell.} quant-ph/0103028;
{\it Proc. Nat. Acad. Sci. USA,} {\bf 98}, 14224 (2001); 
{\it Proc. Nat. Acad. Sci. USA,} {\bf 98}, 14228 (2001); 
{\it Europhys. Lett.}, {\bf 57}, 775 (2002).

[3] R. Gill, G. Weihs, A. Zeilinger, M. Zukowski, {\it Comment on "Exclusion of time in the theorem 
of Bell" by K. Hess and W. Philipp.} quant-ph/0204169 (2002).

[4] I. Pitowsky,  Phys. Rev. Lett, {\bf 48}, N.10, 1299-1302 (1982);
Phys. Rev. D, {\bf 27}, N.10, 2316-2326 (1983);
S.P. Gudder,  J. Math Phys., {\bf 25}, 2397- 2401 (1984);
L. Accardi, {\it Urne e Camaleoni: Dialogo sulla realta,
le leggi del caso e la teoria quantistica.} Il Saggiatore, Rome (1997);
A. Fine,  {\it Phys. Rev. Lett.,} {\bf 48}, 291--295 (1982);
P. Rastal, {\it Found. Phys.,} {\bf 13}, 555 (1983).
W. Muckenheim,  {\it Phys. Rep.,} {\bf 133}, 338--401 (1986);
W. De Baere,  {\it Lett. Nuovo Cimento,} {\bf 39}, 234-238 (1984);
{\bf 25}, 2397- 2401 (1984); 
W. De Muynck, J.T. Stekelenborg,  {\it Annalen der Physik,} {\bf 45},
N.7, 222-234 (1988).

[5] A. Yu. Khrennikov,  {\it Dokl. Akad. Nauk SSSR,} ser. Matem.,
{\bf 322}, No. 6, 1075--1079 (1992); J. Math. Phys., {\bf 32}, No. 4, 932--937 (1991);
{\it Phys. Lett. A,} {\bf 200}, 119--223 (1995);
{\it  J. Math. Phys.,} {\bf 36},
No.12, 6625--6632 (1995); {\it Phys. Lett.},
A, {\bf 278}, 307-314 (2001); {\it J. Math. Phys.}, {\bf 41}, 5934-5944(2000);
{\it Il Nuovo Cimento,} {\bf B 115}, 179-184;
A.Yu. Khrennikov, {\it Non-Archimedean analysis: quantum
paradoxes, dynamical systems and biological models.}
Kluwer Acad.Publ., Dordreht (1997).

[6] Papers of L. Accardy, G. Adenier, W. De Muynck, D. Gross, 
S. Gudder, A. Kracklauer, J.-A. Larsson, I. Volovich in 
{\it Proc. Conf. "Foundations of Probability and Physics",}
editor A. Khrennikov, WSP, Singapore (2001).

[7] Gill R., quant-ph/0110137 (2001).

[8] Khrennikov A. , {\it Interpretations of Probability.}
VSP Int. Sc. Publishers, Utrecht (1999).

[9] H. P. Stapp, {\it Phys. Rev.,} D, {\bf 3}, 1303-1320 (1971);
P.H. Eberhard, {\it Il Nuovo Cimento,} B, {\bf 38}, N.1, 75-80(1977); {\it Phys. Rev. Lett.},
{\bf 49}, 1474-1477 (1982).

[10]  A. Khrennikov, {\it Einstein and Bell, von Mises and Kolmogorov: reality,
locality, frequency and probability.} quant-ph/0006016 (2000).
\end{document}